\documentclass[lettersize,journal]{IEEEtran}
\usepackage{amsmath,amsfonts}
\usepackage{algorithmic}
\usepackage{algorithm}
\usepackage{array}
\usepackage[caption=false,font=normalsize,labelfont=sf,textfont=sf]{subfig}
\usepackage{textcomp}
\usepackage{stfloats}
\usepackage{url}
\usepackage{verbatim}
\usepackage{graphicx}
\usepackage{cite}
\usepackage{booktabs}
\begin{document}

\title{Identifying Privacy Concerns in Upcoming Software Release: A Peek into the Future}

\author{Aurek Chattopadhyay 
    and Nan Niu,~\IEEEmembership{Senior Member, IEEE}
    \thanks{Aurek Chattopadhyay is with the University of Cincinnati, Cincinnati, Ohio, USA (email: chattoak@mail.uc.edu).}
    \thanks{Nan Niu is with the University of North Florida, Jacksonville, Florida, USA (email: nan.niu@unf.edu).}
    }



\maketitle

\begin{abstract}
Identifying the features to be released in the next version of software, from a pool of potential candidates, is a challenging problem. User feedback from app stores is frequently used by software vendors for the evolution of apps across releases. Privacy feedback, although smaller in volume, carries a larger impact influencing app's success. Multiple existing work has focused on summarizing privacy concerns at the app level and has also shown that developers utilize feedback to implement security and privacy-related changes in subsequent releases. However, the current literature offers little support for release managers and developers in identifying privacy concerns prior to release. This gap exists as user reviews are typically available in app stores only after new features of a software system is released. In this paper, we introduce Pre-PI, a novel approach that summarizes privacy concerns for to-be-released features. Our method first maps existing features to semantically similar privacy reviews to learn feature-privacy review relations. We then simulate feedback for candidate features and generate concise summaries of privacy concerns. We evaluate Pre-PI across three real-world apps, and compare it with Hark, a state-of-the-art method that relies on post-release user feedback to identify privacy concerns. Results show that Pre-PI generates additional valid privacy concerns and identifies these concerns earlier than Hark, allowing proactive mitigation prior to release. 

\end{abstract}


\section{Introduction} 
  A key aspect of modern-day software engineering is strategic release planning \cite{Ameller_2017_SANER,Sahin_IEEM_2020,Nayebi_2015_book,Marner_22_Computers} with continuous and rapid release cycles where software updates are pushed out every few weeks \cite{khomn_release_planning,Nayebi_release_planning}.
App stores such as Google Play Store (https://play.google.com/store) have emerged as a way to actively involve users in the continuous development process. User reviews contain important sources of information that can be used by developers for the successful evolution of apps  \cite{scalabrino_2017, NLP_handbook_Maalej_2025}.
While there is an abundance of user feedback available-- meeting the three Vs of big data in terms of volume, velocity, and variety \cite{Gomez_2017_Software}, finding privacy feedback from user reviews is a needle-in-the-haystack problem with less than 0.5\% of reviews being related to security and privacy \cite{Mukherjee20_archiv}.

Although privacy feedback is only a small fraction of user reviews, it still has a great impact. Security and privacy concerns in an app are mostly associated with low star ratings \cite{Mukherjee20_archiv,khalid_2015}. This hurts user trust and late fixes can be expensive. Nyugen et al. \cite{Nguyen_SP_2019} have shown that security and privacy related user reviews play an important role in software release cycles, as developers utilize the feedback to implement security and privacy changes in their app in subsequent releases.
\IEEEpubidadjcol

\begin{figure}[h]
  \centering
  \includegraphics[width=\linewidth]{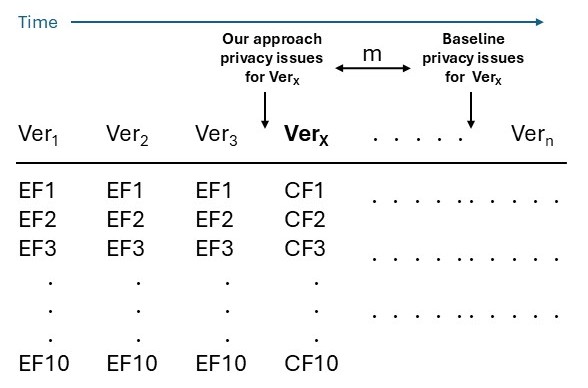}
  \caption{Our approach identifies privacy issues for a release ($Ver_x$) m versions earlier than current existing methods}
  \label{fig:intro_example}
\end{figure}

Existing techniques \cite{ICSE_22_google,ASE_22_Anas,Hark_google} have focused on summarizing privacy concerns from user reviews for developers. However, these approaches are reactive in nature, where they depend on post-release feedback from users that is only received after features are released to the users. As a result, the current literature offers little support for release managers and developers to anticipate privacy concerns before a release occurs, which would allow proactive mitigation. 

This limitation is demonstrated in Figure \ref{fig:intro_example}, where our proposed approach identifies privacy issues for a given release version ($Ver_x$) m versions earlier as compared to existing methods. Each version prior to $Ver_x$ contains features that we refer to as existing feature (EF), which have already been released. Candidate feature (CF) represents the set of features that are part of the release pool for the upcoming version $Ver_x$. Our proposed approach anticipates privacy concerns for the CFs before release. On the other hand, baseline techniques are only able to detect privacy concerns for $Ver_x$ after it is deployed and user feedback becomes available m versions later.

Harkous et al. \cite{Hark_google} have introduced a state-of-the-art technique called Hark, to summarize privacy concerns from user reviews. One of their major contributions is that they use a classifier to extract privacy reviews of an app, and then use generative NLP models to generate a short-concise about 2-4 words summary for each privacy review of the app raised by the users. They call these privacy concerns: ``issues". These issues (e.g., ``Unwanted Password Sharing" and ``Excessive Permissions") act like a glanceable summary for the developers, as they can quickly go through them without having to read the entire privacy reviews. 

In this paper, we propose a novel approach \textit{\textbf{Pre}--release- \textbf{P}rivacy \textbf{I}ssue generation  (\textbf{Pre-PI})} to summarize privacy concerns for to be released features, allowing release managers and developers to receive early feedback and anticipate user reactions before release of the upcoming version of software. 
Motivated by Harkous et al. \cite{Hark_google} we generate glanceable summaries of privacy issues from to be released features. To achieve this we first associate existing features with semantically similar existing reviews to capture privacy signals from past feedback. We then utilize the existing features and user feedback to simulate reviews for the to be released features. This simulated feedback acts as a practical mechanism to fill the feedback gap for to be released feature. The issues are then generated from the simulated user feedback. The differences between our method and the Hark baseline is shown in Figure \ref{fig:privacy_classifier}. While the baseline depends on post release feedback of $Ver_x$ to generate privacy issues, our method allows early issue generation by simulating user reviews for upcoming features. 

We utilize Machine Learning and advanced Natural Language Processing techniques in our approach. The goal of our approach is to provide feature specific privacy feedback before release, so that the release managers can decide which features to release next and developers can act and fix the raised privacy concerns for the features before release. 
The major contributions of our paper are as follows:

\begin{figure*}[t]
  \centering
  \includegraphics[width=\textwidth]{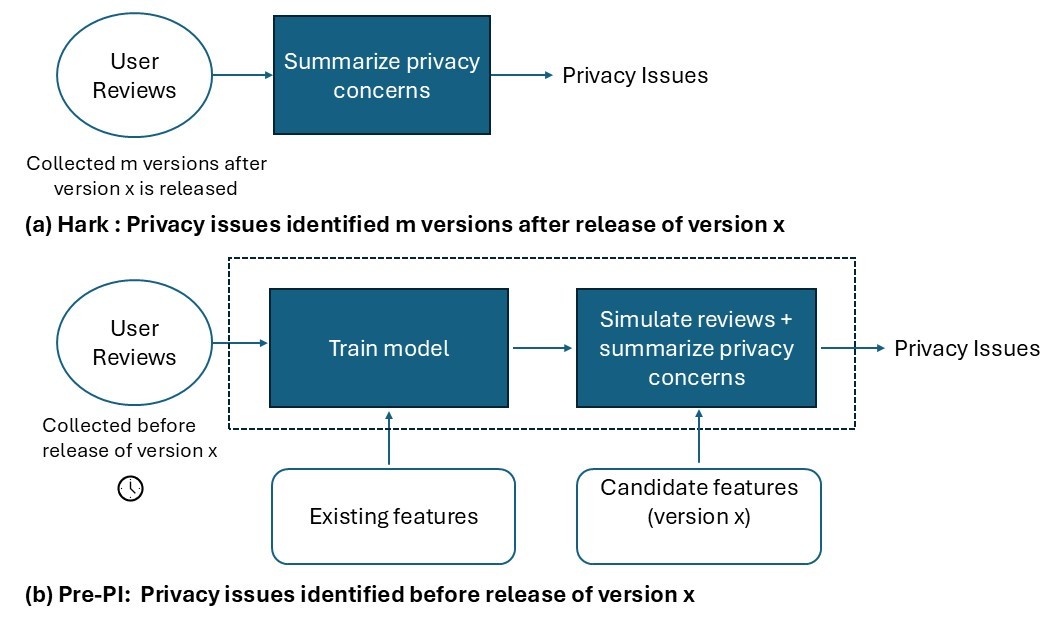} 
  \caption{Comparison of Hark and Pre-PI: Pre-PI identifies privacy issues before release of version x, while Hark depends on user feedback available m versions after release of version x to identify privacy issues.}
  \label{fig:privacy_classifier}
\end{figure*}

\begin{itemize}
    \item {We propose a novel solution Pre-PI that identifies and summarizes privacy concerns for to be released features, enabling proactive privacy concerns evaluation before the upcoming release of software. Our approach generates glancable, structured privacy issues without relying on post-release app feedback. }
    \item {We evaluate our approach across three real world mobile apps by comparing the issues generated by our method  against the baseline Hark. A human subject study involving actual app users shows that the issues identified by our method are valid privacy concerns faced during app usage.}
    \item{ Our method consistently generates a higher number of privacy issues across releases compared to the baseline. Further, we show that privacy issues identified by our method often appear later in user reviews as identfied by the baseline.}
\end{itemize}

The rest of the paper is organized as follows: Section 2 contains the background and related work. Section 3 explains our approach in detail. Section 4 presents the experimental evaluations and reports the results. Section 5 discusses the threats to validity and then we conclude the paper in Section 6.

\section{Background and Related Work}

User reviews are utilized for various activities in modern day requirements engineering (RE) tasks. This is in line with Crowd-based Requirements Engineering (CrowdRE) paradigm, that was first introduced by Groen et al. \cite{REFSQ_15_Groen} as: “a semi-automated RE approach for obtaining and analyzing any kind of user feedback from a crowd, with the goal of deriving validated user requirements.” In this context, the ''crowd” should be considered as a pool of current and potential stakeholders whenever the users through user feedback share their feedback and experiences with a software product \cite{REFSQ_15_Groen}.  Empirical studies \cite{ICSME_15_Palomba,JSS_18_Palomba} have shown that developers use reviews collected from crowd-sourced apps to make changes to their code. Furthermore, numerous studies \cite{scalabrino_2017,ICSE_17_Palomba,REW_23_Stronstad,RE_23_Nayebi} have explored the utilization of user reviews from app stores to address the various aspects of release planning activities.

Among the various types of user feedback, privacy feedback although low in number, carry a big impact. Khalid et al. \cite{khalid_2015} conducted a study of over 6,000 low star rated user reviews across 20 iOS apps and identified 12 types of user complaints. They found that privacy and ethics complaints were among the most negatively perceived by users, and had a more damaging impact on the app reputation than other complaints such as app crashing or feature removal. Besmer et al. \cite{besmer_2020} further observed that privacy app reviews are associated with lower star ratings and more negative sentiment compared to non privacy reviews, yet privacy reviews recieve higher engagement as measured through helpful and unhelpful votes.

Importantly, privacy feedback is not just expressive, but acts as actionable signals for developers and release managers. Nguyen et al. \cite{Nguyen_SP_2019} studied the relationship between security and privacy related reviews (SPR) and corresponding security and privacy related app updates (SPU). Analyzing over 4.5 million user reviews and 62,000 app versions from 2,582 apps on the Google play store, a SVM review classifier was trained to identify SPRs and static code analysis was used to detect privacy related app updates in subsequent versions of the app. They found that SPRs were mapped to SPUs in 60.77\% of all cases, showing that user feedback on security and privacy related concerns lead to actual privacy improvements by software teams. Nguyen et al. \cite{Nguyen_SP_2019} highlight that SPR are significant predictors for SPU and in majority of cases the developers change the respective app code or respond to the users justifying why certain permissions are required. 

\subsection{Summarizing Privacy Concerns}
Even though privacy feedback has a high impact, it remains scattered and sparse across the large volume of user feedback. Maleej et al.\cite{NLP_handbook_Maalej_2025} identified two main challenges that software vendors face in achieving the full potential of user feedback: (1) there is a large quantity of feedback data that is hard to manage manually; (2) the inconsistent quality of feedback, as feedback from users can be uninformative, repetitive, or inaccurate.

Summarizing privacy concerns from user feedback is a scalable way of gaining a high-level overview of the concerns raised by users. Summarization techniques generally fall into two categories: extractive summarization and abstractive summarization \cite{Mehta_2019_summarization}. 
The extractive summarization technique extracts the most representative sentences directly from the original text, whereas the abstractive summarization technique captures the main idea and may include novel sentences not present in the original text \cite{NLP_handbook_Maalej_2025}.
Prior work has explored both approaches in the context of privacy concerns. Extractive summarization has been explored by Ebrahimi and Mahmoud \cite{ASE_22_Anas} and Nema et al. \cite{ICSE_22_google}, while abstrative summarization was proposed by Harkous et al.  \cite{Hark_google}. We discuss a more detailed overview of these approaches next. 

Ebrahimi and Mahmoud \cite{ASE_22_Anas} proposed an extractive summarization technique for privacy concerns in mobile apps. They observed that privacy feedback is both sparse and domain-dependent, making it challenging to capture using generic data mining techniques. Their method first involves a manual analysis of one and two star rated user reviews to extract domain-specific privacy indicator keywords across three app domains investing, mental health, and food delivery apps. The second part of their approach uses the domain-specific privacy indicator keywords to guide an extractive summarization algorithm where hybrid TF-IDF and Glove techniques are used to score and select the relevant reviews. Therefore, generating privacy summaries for each domain. 

Nema et al. \cite{ICSE_22_google} conducted a large-scale analysis of privacy concerns in mobile app reviews. They processed reviews using Natural Language Understanding components as opposed to traditional approaches such as those relying on bag-of-words representation. Their motivation came from the limitations of traditional approaches that fail to take the sequence of words into account and cannot differentiate between  the reviews containing the same words in different order such as ``delete cookies and website history" and ``my article on history of cookies got deleted from website" \cite{ICSE_22_google}. They developed an ensemble classifier combining three deep learning based NLP models: Universal Sentence Encoder (USE), Bidirectional Encoder Representations from Transformers (BERT), and Sentiment-aware BERT (BERT-SST) to train a privacy classifier. After classification of the privacy reviews, they applied k-means clustering algorithm for summarization to group similar reviews together. They ranked the reviews within each cluster using silhouette scores and used the top ten reviews with the highest scores as representative reviews for each cluster. Further, for a subset of clusters they manually reviewed the top ranked twenty reviews for each representative cluster to assign a short thematic label. The dominant themes identified by them were ``too many permissions" and ``too much personal information", whereas concerns such as ``privacy controls", ``tracking" and ``selling data" were observed across some of the app categories. 

Further complementing these findings, Xie et al. \cite{ICSE_24_Xie} studied the excessive data collection in virtual private assistants such as Amazon Alexa. They proposed a method that detects inconsistencies by identifying mismatches between the data requested by the app and the data required for the functionality of the app. Their study revealed that 21.7\% of analyzed apps from the Alexa app store demonstrated suspicious data collection behaviors. 

Harkous et al. \cite{Hark_google} introduced an abstractive summarization method for privacy related user feedback. They developed a system called Hark aimed at imrproving user to developer communication, where developers can have access to glanceable summaries of the privacy concerns, instead of having to go through all the privacy reviews. They propose three main requirements that would help developers go through the privacy reviews in an efficient manner: (1) High coverage of the privacy concerns, irrespective of the way they are linguistically expressed in the reviews. (2) Gist of the concerns raised, which would enable developers to understand the context without having to go through all the reviews. (3) It should enable developers to have a higher level understanding with the option of diving deeper into the issues. To meet these requirements, Harkous et al. \cite{Hark_google} introduced a privacy review classifier for the first requirement, an issue generation model for the second requirement, and a theme generation component for the third requirement. 

Hark utilizes state-of-the-art NLP techniques such as the generative T5 model by Raffel et al. \cite{Raffel_T5}. The privacy review classifier was trained on a manually annotated dataset using the T5 model, where it serves as a privacy review classification model. The issue generation model also based on T5, generates abstractive summaries from the reviews. Each privacy review would generate one or more issues. The issue generation model typically generates short 2-4 word actionable phrases referred to as issues. Some examples include \textit{Unwanted Password Sharing, Excessive Permissions,} and \textit{Personal Address Deletion}. These issues would highlight the actual topics discussed by the user rather than high-level concepts such as bugs or feature requests \cite{Hark_google}. To train the issue generation model, Harkous et al. \cite{Hark_google} sampled 1,060 reviews from their original dataset and manually annotated these with the set of issues for each review. Finally, the theme generation component of Hark clusters similar issues together into higher level privacy themes, supporting a high-level overview, while allowing developers to have the option of diving deep into the individual issues. 

We are inspired from the work of Harkous et al. \cite{Hark_google} specifically the issue generation model that provides a gist of privacy concerns raised in the user reviews as a summary. Our approach in this paper further builds upon this by moving on from being a reactive system to a pre-release privacy feedback assessment system. Motivated by Nguyen et al. \cite{Nguyen_SP_2019} we map features to reviews and propose a two-stage issue generation model that first learns existing feature-review associations, and then generates issues for candidate features. This helps to generate issues for upcoming release with the help of candidate features in the absence of actual user reviews, thereby helping release managers to evaluate feature-level privacy risks early and make informed decisions on candidate feature release. Additionally, our approach also helps developers proactively address or clarify potential issues before releasing the upcoming version of software.

\section{Methodology}
\begin{figure*}[t]
  \centering
  \includegraphics[width=\textwidth]{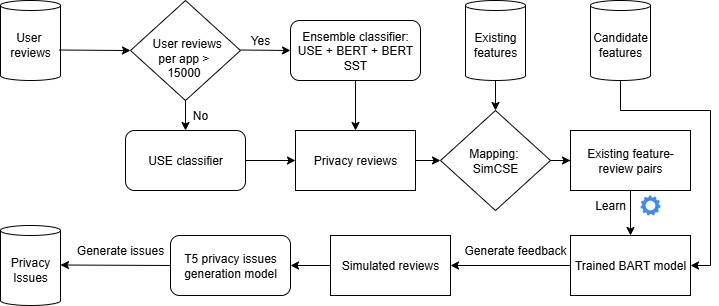} 
  \caption{Overview of our proposed approach: Pre-PI}
  \label{fig:method_figure}
\end{figure*}

An overview of our approach Pre-PI is shown in Figure \ref{fig:method_figure}. In this section, we discuss the detailed steps and implementation of our approach.

\textbf{Step 1: Privacy classification of user reviews:} The first step of our approach is classifying reviews as privacy or not. To acheive this, we utilize the dataset shared by Ebrahimi et al. \cite{ASE_22_Anas}, which contains 8,648 manually labeled privacy and non-privacy reviews. Our privacy classifier is inspired by the multi model ensemble classifier proposed by Nema et al. \cite{ICSE_22_google}. Specifically we train three state-of-the-art deep learning based language models proposed by Nema et al.  \cite{ICSE_22_google}. We utilized the same architecture and training parameters based on their methodology. These three models are explained next. \textbf{(i) BERT model:} We utilized the pre-trained BERT model and added an additional layer after the last layer for classification into privacy and non privacy classes. The BERT model is fine-tuned for three epochs on the dataset, and then we select the best epoch model based on validation performance. \textbf{(ii) Sentiment-aware BERT (BERT-SST):} As there is a strong correlation between privacy reviews and negative sentiment, following Nema et al. \cite{ICSE_22_google}, the BERT model was first fine-tuned on Stanford Sentiment Treebank 2 dataset to make it sentiment-aware. We then fine-tune the model on the privacy dataset for three epochs and then select the best model based on performance in validation set.
\textbf{(iii) Universal Sentence Encoder (USE) Model:}  We utilized the pre-trained USE model to generate 512 dimensional sentence embeddings capturing the semantic content of reviews without a need for fine-tuning. This embedding is then fed through a two layered feed forward neural network with 512 hidden units per layer. This model is trained for 20 epochs and  we select the best epoch model based on validation set performance. The performance metrics of these three classifiers is shown in Table~\ref{tab:privacy_classifier_metrics}. We adopt the ensemble classification model introduced by Nema et al. \cite{ICSE_22_google} where a review is selected as privacy, only if all the three models-- BERT, BERT-SST, and USE predict the review as privacy. This conservative model prioritizes precision at the cost of recall and was identified by Nema et al. \cite{ICSE_22_google} as the best performing model. Motivated by their findings, we apply the ensemble model for apps containing more than 15000 reviews, where sufficient data allows for stricter filtering. On the other hand, for apps with fewer than 15000 reviews, we relax the criteria and utilize the USE classifier, as it achieved highest precision compared to the other two classifiers.

\begin{table}[h]
\centering
\caption{Performance metrics of the privacy classifiers}
\label{tab:privacy_classifier_metrics}
\begin{tabular}{|l|c|c|c|}
\hline
\textbf{Metrics}   & \textbf{BERT} & \textbf{BERT-SST} & \textbf{USE} \\
\hline
\textbf{Accuracy}  & 0.91            & 0.92                & 0.90           \\
\textbf{Precision} & 0.81            & 0.79                & 0.82           \\
\textbf{Recall}    & 0.81            & 0.90                & 0.76           \\
\textbf{F1 score}  & 0.81            & 0.84                & 0.79           \\
\textbf{AUC}       & 0.96            & 0.96                & 0.95           \\
\hline
\end{tabular}
\end{table}

\textbf{Step 2: Mapping privacy reviews to features:} To associate privacy related feedback with app features, we leverage Simple Contrastive Learning of Sentence Embeddings (SimCSE), a state-of-the-art technique introduced by Gao et al. \cite{simcse}. SimCSE leverages contrastive learning to generate high quality sentence embeddings, which are very effective for semantic textual similarity tasks. We utilize the pre-trained SimCSE model to generate embeddings for both features and privacy reviews. For each feature, cosine similarity is computed between its embedding and the embeddings of all the privacy reviews. The privacy reviews are then ranked based on similarity, and the top ten most semantically similar privacy reviews are selected and matched to that particular feature. 

\textbf{Step 3: Simulated reviews model training:} To simulate privacy-related feedback for candidate features, a dataset of feature-review pairs was created, where each feature is paired with its top ten most semantically similar privacy reviews (as identified in step 2). These pairs were then used to fine tune a Bidirectional and Auto-Regressive Transformers (BART) model \cite{bart}. BART is a transformer-based architecture that combines the strengths of both BERT and GPT \cite{bart}. The feature-review pairs dataset was split into a 90:10 ratio of training and validation set and the BART model was fine-tuned for 5 epochs with early stopping enabled, to prevent the model from overfitting. BART model is shown to be very effective when fine-tuned for text generation and understanding tasks \cite{bart}, making it highly suitable for the simulated reviews generation step. We selected BART over T5 as BART understands the context better and provides a richer summary \cite{gaby_bart_t5}. 

\textbf{Step 4: Issue generation model:} To create a model that would generate concise summaries of privacy concerns, we randomly sampled 500 privacy reviews from the reviews shared by Ebrahimi et al. \cite{ASE_22_Anas} and manually annotated these reviews with a short descriptive summary of privacy issues. The issue labeling process was similar to that of Harkous et al. \cite{Hark_google}, where they suggested issues to be about 2-4 words highlighting the actual topic raised by users. Each review was labeled with one or more issues. Inspired from Harkous et al. \cite{Hark_google} these review-issue pairs were then used to fine-tune a T5 model for abstractive issue generation. We performed a 90-10 percent train-test split and trained the model 5 epochs on the T5 base model. We selected the T5 base architecture rather than the larger T5-11B model used by Harkous et al. \cite{Hark_google}, due to the smaller size of our dataset and to prevent overfitting. We selected a learning rate of 0.005 and label smoothing coefficient of 0.1 consistent with Hark parameters. 

\textbf{Step 5: Synthetic model inference and issue generation:} For each candidate feature, we utilized the trained BART model from step 3 to generate ten simulated privacy reviews. These reviews simulate the kind of feedback that might be raised after the feature is released. These generated reviews were then forwarded to the trained T5 issue generation model from step 4 to generate one or more privacy issues. This two-stage pipeline provides us with privacy issues associated with the candidate features, allowing pre-release privacy feedback assessment. 

\section{Experimental Evaluation}
In this section, we evaluate the performance of our proposed approach Pre-PI, against the baseline Hark. First, we begin by summarizing some of the key implementation details of the baseline Hark. This is followed by the description of the datasets used for evaluation. We then introduce the research questions and present the results and analysis. 

\subsection{Baseline: Hark}

We evaluate the performance of Pre-PI against Hark \cite{Hark_google} that serves as a baseline. To ensure a fair and valid comparison, we make certain changes to our implementation of Hark. The major operational differences along with the rationale behind them is explained next. We also share our replication of Hark baseline in the replication packet (https://doi.org/10.5281/zenodo.17274646).

In Hark, a T5 model is trained for classification of reviews into privacy and non-privacy classes. Since T5 is a generative model and more suitable for generative tasks, we refrained from using it for classification tasks, and instead we utilized the same multi-model ensemble classifier which we used for Pre-PI. This choice further ensured consistency among Pre-PI and Hark enabling a fair comparison.
As the datasets used by Harkous et al. \cite{Hark_google} were not publicly available, we relied on the same training samples for the issue generation model, as Pre-PI. To further ensure a valid and fair comparison, we used the T5 base architecture similar to Pre-PI rather than the larger T5-11B model used by Harkous et al. \cite{Hark_google}. This model is more suitable for our smaller datasets and mitigates the risk of overfitting. Another difference lies in the formulation of issue sets. Hark generated privacy issues at the app level, without associating them to specific software releases. For our evaluation setup, we adapted the output of Hark by partitioning the issues into seperate release specific feedback sets based on the timestamps of the corresponding user reviews.

\subsection{Datasets}
To evaluate our approach, we utilized three app datasets from the real world: Zoom, Microsoft 365 Word, and Webex. New and enhanced features are frequently found in software release notes \cite{Bi_TSE_22,Moreno_TSE_17}. We extracted features directly from publicly available release notes provided by the software vendors. We also collected the date of release of these extracted features from the release notes. To collect app reviews, we extracted them from user reviews posted on the Google Play store. We define a release instance as the set of features released within a time period. For example, all features released in February 2025 are grouped together as one release instance. Two consecutive release instances are shown in Table \ref{table:Instance}. The top 10 features released in January 2025 are part of one release instance, while the 19 features released in February 2025 become a part of another release instance. 

For each app, we divide the release instances into two groups: the existing feature group and the candidate feature group. The existing feature group consists of the earlier release instances and contains the features that were used to train the models. The candidate feature group consists of the more recent instances and was reserved for evaluation. For example, in the Microsoft 365 Word dataset, the existing feature group contains releases from 2018 January to 2023 December, while the candidate feature group covers releases from 2024 January to 2024 December.

The user reviews were extracted from a Google Play Store scrapper \cite{Google-Play-Scraper-Web}. We extracted user reviews for the same time period as the release instances. The existing features were aligned with all the user feedback during their time period, while for the candidate features group, the user reviews were aligned with each corresponding release instance. 

The characteristics of the datasets are summarized in Table \ref{table:dataset}. For each app, the table reports the number of existing features, the time period of the existing features, and the number of privacy feedback messages associated with them. The table also shows the number of candidate features, the time period of those candidate features, and the number of privacy messages corresponding to them. Therefore, the table demonstrates the dataset characteristics of all the three apps with  details of the existing features and candidate features data.

\begin{table*}[!t]
\centering
\caption{Two consecutive release instances of the Zoom dataset (the top 10 features released in January 2025 are part of one instance and the bottom 19 features released in February 2025 are part of another instance).}\label{table:Instance}
\normalsize
\begin{tabular}{ | r | l | c |}
\hline
\multicolumn{1}{|c}{\bf \#} & \multicolumn{1}{|c|}{\bf Feature Description} & $\!\!${\bf Release Month} \\
\hline \hline
1 & Users can send contact requests with a prefilled message during Zoom meetings through & Jan 25\\
& $\qquad$ the Profile Card. When sending a contact request, the message field automatically   &\\
& $\qquad$ includes contextual information about the current meeting, including the meeting   &\\
& $\qquad$ name and date. This prefilled message helps recipients understand the context of the   &\\
& $\qquad$ connection request, making it easier to identify and accept contact requests from &\\
& $\qquad$  meeting participants.&\\
2 & Hosts can schedule webinars to start with a pre-recorded video and automatically roll  & Jan 25\\
& $\qquad$ over to a live session after the recording completes. Hosts have the option to   &\\
& $\qquad$ end the simulive portion early and transition to live manually. Attendees experience     &\\
& $\qquad$ a seamless transition from the pre-recorded content to the live segment. &\\ 
& $\qquad$ Hosts and panelists are notified before the rollover to prepare for the live portion. &\\
& $\qquad$ This feature ensures webinars start on time with engaging content while enabling   &\\
& $\qquad$ live interaction with the audience. Users must be on Zoom Workplace App version &\\
& $\qquad$ 6.3.5 or later to use this feature.&\\
& \ldots $\qquad\qquad$ \ldots $\qquad\qquad$ \ldots $\qquad\qquad$ \ldots $\qquad\qquad$ \ldots $\qquad\qquad$ \ldots $\qquad\qquad$ \ldots $\qquad\qquad$ \ldots & \ldots \\
10 & Account owners and admins can set channel names and descriptions up to 64 characters  & Jan 25\\
& $\qquad$for Zoom Push to Talk. This enhancement enables more descriptive and   &\\
& $\qquad$meaningful channel identifiers. The increased length is configurable through the   &\\
& $\qquad$admin web portal and API, though the Zoom app will not reflect the full name.  &\\
& $\qquad$Admins will be informed to adopt a naming convention as the Zoom app may have &\\
& $\qquad$limited display space.&\\
\hline
1 & Meeting participants can request access to recordings directly from the meeting card or & Feb 25 \\
& $\qquad$ recording link without having to send separate messages through chat, email, or  &\\
& $\qquad$ calls. Meeting hosts receive these requests through in-product notifications &\\
& $\qquad$  and emails, which direct them to a viewing page where they can manage access  &\\
& $\qquad$ permissions. Hosts can view all requesters on the share modal, grant or deny access,&\\
& $\qquad$ and manage existing permissions for specific users. The feature respects all security & \\
& $\qquad$ settings, including authentication requirements and password protection.  & Feb 25 \\
2 & After a meeting ends, a dynamic pop-up appears, based on the assets utilized during   & \\
& $\qquad$the meeting, directing users to the meeting details page. Here, they can access & \\
& $\qquad$available meeting assets such as meeting recording, summary, continuous chat log,  &\\
& $\qquad$ as well as any content shared during the session, such as whiteboards, links, and notes. & \\
& \ldots $\qquad\qquad$ \ldots $\qquad\qquad$ \ldots $\qquad\qquad$ \ldots $\qquad\qquad$ \ldots $\qquad\qquad$ \ldots $\qquad\qquad$ \ldots $\qquad\qquad$ \ldots & \ldots \\
19 & Users can view a complete list of meeting participants directly within the primary meeting  & Feb 25 \\
& $\qquad$card interface. The participant list appears in the full meeting page, providing clear & \\
& $\qquad$visibility of all attendees. & \\

\hline
\end{tabular}
\normalsize
\end{table*}

\begin{table*}[]
\centering
\caption{Characteristics of Datasets}\label{table:dataset}
\begin{tabular}{|@{}l|lll|lll@{}|}
\toprule
\textbf{\begin{tabular}[c]{@{}l@{}}Dataset {[}source of\\ release notes{]}\end{tabular}} & \textbf{\begin{tabular}[c]{@{}l@{}}\# of existing \\ features\end{tabular}} & \textbf{\begin{tabular}[c]{@{}l@{}}Existing features\\ time-period\end{tabular}} & \textbf{\begin{tabular}[c]{@{}l@{}}\# of privacy \\ feedback \\ messages \\ for existing \\ features\end{tabular}} & \textbf{\begin{tabular}[c]{@{}l@{}}\# of candidate \\ features\end{tabular}} & \textbf{\begin{tabular}[c]{@{}l@{}}Candidate features\\ time-period\end{tabular}} & \textbf{\begin{tabular}[c]{@{}l@{}}\# of privacy \\ feedback\\ messages for \\ candidate\\ features\end{tabular}} \\ \midrule
\textbf{Webex \cite{webex}} & 219 & 2022 Jan - 2023 May & 923 & 145 & 2023 June - 2024 June & 302 \\
\textbf{Zoom \cite{zoom}} & 640 & 2022 Jan - 2024 April & 340 & 273 & 2024 May - 2025 Feb & 63 \\
\textbf{\begin{tabular}[c]{@{}l@{}}Microsoft 365 \\ Word {\cite{word}}\end{tabular}} & 203 & 2018 Jan - 2023 Dec & 969 & 99 & 2024 Jan - 2024 Dec & 76 \\ \bottomrule
\end{tabular}
\end{table*}

\subsection{Research Questions (RQs): Definition and Design}
We answer the following RQs:

\textbf{RQ1: Do the issues generated by Pre-PI reflect actual privacy concerns faced by users during app usage and how do they compare against the issues generated by baseline?}

To answer RQ1, we conducted a human subject study involving two evaluators for each of the apps, each having at least 5 years of experience as a user of that particular app. Informed consent was obtained from the evaluators. To prevent evaluator fatigue, we provided each evaluator with a subset of issues generated by both Pre-PI and Hark for the particular app being evaluated. To ensure fairness, the method utilized to generate the issue was not provided to the evaluators. For each issue, the evaluators were asked to indicate whether it represented a valid privacy concern for the app, based on their experience as users. 

\textbf{RQ2: How do the issues that overlap between Pre-PI and baseline evolve across app releases over time?}

To address RQ2, we compare the sets of overlapping issues identified by Pre-PI and Hark for successive app releases. While Hark extracts issues from actual user reviews, Pre-PI generates issues only depending on candidate features without relying on post-release user feedback. We compute the overlap ratio for each method as follows:

\[
\text{(i) Hark Overlap Ratio}_i = \frac{|\bigcup_{j=0}^{i} \text{Overlapping Issues}_j|}{|\bigcup_{j=0}^{i} \text{Hark Issues}_j|}
\]

\[
\text{(ii) Pre-PI Overlap Ratio}_i = \frac{|\bigcup_{j=0}^{i} \text{Overlapping Issues}_j|}{|\bigcup_{j=0}^{i} \text{Pre-PI Issues}_j|}
\]

\vspace{0.2cm}

\noindent where i represents the candidate feature release version ranging from 0 to n, with i=0 representing $Ver_x$.
We compute the overlap ratios cumulatively for each release. For a given release i, we consider the union of all the issues generated from release j=0 through j=i. So, all issues from release 0 to release i are included. For Pre-PI we assume user feedback is available only prior to release 0 and Pre-PI generates privacy issues solely relying on candidate features from release 0. These ratios help us evaluate the extent to which the two methods align with the overlapping issues set, and whether Pre-PI is able to proactively identify the issues that are identified from the Hark baseline through user feedback.

\textbf{RQ3: Does Pre-PI identify issues earlier than the baseline method?}

In RQ3, we factor in the temporal emergence of privacy issues between Pre-PI and Hark. Specifically, for a given release i, if Pre-PI generates a set of issues that are not present in the baseline Hark output up to version i, but are subsequently identified by Hark within a defined evaluation window (n), we consider those as valid early predictions. This represents the issues that the baseline initially failed to detect, but was identified by Pre-PI early. To reflect this, we extend the overlap ratio introduced in RQ2 and define \textbf{\textit{temporal overlap ratio}}, where for a given candidate feature release i, the overlapping set includes (i) cumulative issues commonly identified by both methods up to release i. (ii) cumulative issues identified by Pre-PI up to release i and later confirmed by Hark within the evaluation window (n) as valid. This adjustment enables us to further strengthen the evaluation towards the anticipatory power of Pre-PI.

\subsection{Results and Analysis}
To answer RQ1, we summarize the results of the evaluator scores of the human subject study for the three apps in Table \ref{table:RQ1}. The table shows the percentage of issues selected by each evaluator as valid across three categories: common issues, unique Hark issues, and unique Pre-PI issues. Common issues refer to the issues that were identified by both the methods: Hark and Pre-PI. Unique Hark issues are the set of issues that only Hark was able to identify, while unique Pre-PI issues are the set of issues that were only identified by Pre-PI and was not identified by Hark. We show the evaluator 1, evaluator 2, and overall percentages across the three categories. 

Across all apps, the evaluators consistently selected common issues as valid with mean percentage rates ranging from 0.75 to 0.88, indicating that overlap between these two methods captures valid privacy concerns. In contrast, unique Hark issues received lower mean validity rates: 0.56 for Webex, 0.61 for Zoom, and 0.53 for Word, suggesting that many of the unique Hark issues are less relevant. In comparison, unique Pre-PI issues were selected as valid issues by the evaluators at much higher rates: 0.91 for Webex, 0.78 for Zoom, and 0.73 for Word. These results indicate that Pre-PI introduces additional valid issues that evaluators agree to be valid privacy concerns.

Overall, these results show that Pre-PI contributes higher proportions of valid issues in comparison to Hark. Additionally, the common issues established among both methods are valid privacy concerns. In the absence of a ground truth, these results motivate us to utilize the overlap of common issues between the methods, as a measure of performance for RQ2 and RQ3.  

To further complement to the RQ1 results, we measured the inter-rater agreement using Cohen Kappa statistic \cite{Cohen_kappa} and percentage of agreement between the evaluators, reported in Table \ref{table:RQ1 cohen kappa}. The overall Cohen kappa score ranges from 0.22 to 0.51, indicating fair to moderate agreement. In addition, the overall percentage agreement values are high. Unique Pre-PI issues achieve higher kappa and percentage agreement values for Zoom and Word, whereas Kappa values for Webex are low. The kappa values and percentage agreement values for unique Hark issues are generally low across all apps, demonstrating that evaluators disagreed more frequently on the validity of unique Hark issues. Common issues achieve good percentage agreement values, although the cohen kappa score varies across apps. 

\begin{table*}[]
\caption{RQ1 results}\label{table:RQ1}
\resizebox{2\columnwidth}{!}{%
\begin{tabular}{@{}c|ccc|ccc|ccc@{}}
\toprule
 & \multicolumn{3}{c}{\textbf{Webex}} & \multicolumn{3}{c}{\textbf{Zoom}} & \multicolumn{3}{c}{\textbf{Word}} \\ \midrule
 & \textbf{\begin{tabular}[c]{@{}c@{}}Evaluator 1 \\ percentage\end{tabular}} & \textbf{\begin{tabular}[c]{@{}c@{}}Evaluator 2 \\ percentage\end{tabular}} & \textbf{\begin{tabular}[c]{@{}c@{}}Mean \\ percentage\end{tabular}} & \textbf{\begin{tabular}[c]{@{}c@{}}Evaluator 1 \\ percentage\end{tabular}} & \textbf{\begin{tabular}[c]{@{}c@{}}Evaluator 2 \\ percentage\end{tabular}} & \textbf{\begin{tabular}[c]{@{}c@{}}Mean\\ percentage\end{tabular}} & \textbf{\begin{tabular}[c]{@{}c@{}}Evaluator 1 \\ percentage\end{tabular}} & \textbf{\begin{tabular}[c] {@{}c@{}}Evaluator 2 \\ percentage\end{tabular}} & \textbf{\begin{tabular}[c]{@{}c@{}}Mean\\ percentage\end{tabular}} \\
\textbf{Common issues} & 0.75 & 1 & 0.88 & 0.8 & 0.85 & 0.83 & 0.73 & 0.77 & 0.75 \\
\textbf{Unique Hark issues} & 0.35 & 0.76 & 0.56 & 0.48 & 0.74 & 0.61 & 0.39 & 0.67 & 0.53 \\
\textbf{Unique Pre-PI issues} & 0.81 & 1 & 0.91 & 0.73 & 0.83 & 0.78 & 0.71 & 0.74 & 0.73 \\ \bottomrule
\end{tabular}%
}
\end{table*}

\begin{table*}[]
\caption{RQ1: Cohen kappa scores}\label{table:RQ1 cohen kappa}
\resizebox{2\columnwidth}{!}{%
\begin{tabular}{@{}ccc|ccc|ccc@{}}
\toprule
\multicolumn{3}{c}{\textbf{Webex}} & \multicolumn{3}{c}{\textbf{Zoom}} & \multicolumn{3}{c}{\textbf{Word}} \\   \midrule
 & \textbf{\begin{tabular}[c]{@{}c@{}}Cohen kappa \\ score\end{tabular}} & \textbf{\begin{tabular}[c]{@{}c@{}}Percentage \\ agreement\end{tabular}} &  & \textbf{\begin{tabular}[c]{@{}c@{}}Cohen kappa \\ score\end{tabular}} & \textbf{\begin{tabular}[c]{@{}c@{}}Percentage \\ agreement\end{tabular}} &  & \textbf{\begin{tabular}[c]{@{}c@{}}Cohen kappa \\ score\end{tabular}} & \textbf{\begin{tabular}[c]{@{}c@{}}Percentage \\ agreement\end{tabular}} \\  \midrule
\textbf{Overall} & 0.22 & 0.67 & \textbf{Overall} & 0.31 & 0.73 & \textbf{Overall} & 0.51 & 0.78 \\
\textbf{Common issues} & 0 & 0.75 & \textbf{Common Issues} & 0.14 & 0.75 & \textbf{Common issues} & 0.56 & 0.83 \\
\textbf{Hark unique issues} & 0.19 & 0.53 & \textbf{Hark unique issues} & -0.02 & 0.48 & \textbf{Hark unique issues} & 0.15 & 0.55 \\
\textbf{Pre-PI unique issues} & 0 & 0.81 & \textbf{Pre-PI unique issues} & 0.71 & 0.9 & \textbf{Pre-PI unique issues} & 0.92 & 0.97
\\
\bottomrule
\end{tabular}%
}
\end{table*}

To address RQ2, we present the performances of Hark and Pre-PI across the three apps: Word, Zoom, and Webex in Tables \ref{table:RQ2 Word}, \ref{table:RQ2 Zoom}, and \ref{table:RQ2 Webex}.  For each app, we report the instance-wise cumulative results. Specifically, for each candidate feature instance, we show the number of issues generated by Hark, the number of issues generated by Pre-PI, the issues that are common to both sets, and the number of unique issues for each method. Based on these values, we compute the cumulative overlap ratios for Hark and Pre-PI. 

Analyzing the growth of identified issues, we observe that Pre-PI shows a clear head start across all the apps. For Word, Pre-PI generates more issues than Hark across most instances and converges to a stable set of issues earlier, demonstrating its strength in finding issues proactively. Similarly for Zoom, Pre-PI consistently generates a higher number of issues and converges sooner compared to the baseline. For Webex, Hark eventually generates more number of issues, however Pre-PI converges faster highlighting its advantage in early detection of privacy concerns.

\begin{table}[]
\centering
\caption{RQ2 Word results}\label{table:RQ2 Word}
\resizebox{\columnwidth}{!}{%
\begin{tabular}{llllllll}
\toprule
\multicolumn{8}{c}{\textbf{Word}} \\
\midrule
\textbf{Instance \#} & \textbf{\begin{tabular}[c]{@{}l@{}}Hark \\ issues\end{tabular}} & \textbf{\begin{tabular}[c]{@{}l@{}}Pre-PI \\ issues\end{tabular}} & \textbf{\begin{tabular}[c]{@{}l@{}}Common \\ issues\end{tabular}} & \textbf{\begin{tabular}[c]{@{}l@{}}Unique \\ Hark \\ issues\end{tabular}} & \textbf{\begin{tabular}[c]{@{}l@{}}Unique \\ Pre-PI \\ issues\end{tabular}} & \textbf{\begin{tabular}[c]{@{}l@{}}Hark \\ overlap \\ ratio\end{tabular}} & \textbf{\begin{tabular}[c]{@{}l@{}}Pre-PI \\ overlap \\ ratio\end{tabular}} \\
\midrule
\textbf{Instance 1} & 5 & 8 & 0 & 5 & 8 & 0 & 0 \\
\textbf{Instance 2} & 14 & 28 & 6 & 8 & 22 & 0.43 & 0.21  \\
\textbf{Instance 3} & 27 & 39 & 7 & 20 & 32 & 0.26 & 0.18  \\
\textbf{Instance 4} & 32 & 40 & 9 & 23 & 31 & 0.28 & 0.23  \\
\textbf{Instance 5} & 39 & 42 & 11 & 28 & 31 & 0.28 & 0.26 \\
\textbf{Instance 6} & 43 & 50 & 18 & 25 & 32 & 0.42 & 0.36 \\
\textbf{Instance 7} & 46 & 52 & 18 & 28 & 34 & 0.39 & 0.35  \\
\textbf{Instance  8} & 49 & 53 & 19 & 30 & 34 & 0.39 & 0.36  \\
\textbf{Instance  9} & 53 & 58 & 24 & 29 & 34 & 0.45 & 0.41 \\
\textbf{Instance 10} & 58 & 61 & 26 & 32 & 35 & 0.45 & 0.43  \\
\textbf{Instance 11} & 63 & 61 & 30 & 33 & 31 & 0.48 & 0.49 \\
\textbf{Instance 12} & 63 & 61 & 30 & 33 & 31 & 0.48 & 0.49 \\
\bottomrule
\end{tabular}%
}
\end{table}

\begin{table}[]
\centering
\caption{RQ2 Zoom results}\label{table:RQ2 Zoom}
\resizebox{\columnwidth}{!}{%
\begin{tabular}{llllllll}
\toprule
\multicolumn{8}{c}{\textbf{Zoom}} \\
\midrule
\textbf{Instance \#} & \textbf{\begin{tabular}[c]{@{}l@{}}Hark \\ issues\end{tabular}} & \textbf{\begin{tabular}[c]{@{}l@{}}Pre-PI \\ issues\end{tabular}} & \textbf{\begin{tabular}[c]{@{}l@{}}Common \\ issues\end{tabular}} & \textbf{\begin{tabular}[c]{@{}l@{}}Unique \\ Hark \\ issues\end{tabular}} & \textbf{\begin{tabular}[c]{@{}l@{}}Unique \\ Pre-PI \\ issues\end{tabular}} & \textbf{\begin{tabular}[c]{@{}l@{}}Hark \\ overlap \\ ratio\end{tabular}} & \textbf{\begin{tabular}[c]{@{}l@{}}Pre-PI \\ overlap \\ ratio\end{tabular}} \\
\midrule
\textbf{Instance 1} & 5 & 33 & 3 & 2 & 30 & 0.6 & 0.09 \\
\textbf{Instance 2} & 12 & 40 & 7 & 5 & 33 & 0.58 & 0.18 \\
\textbf{Instance 3} & 19 & 42 & 10 & 9 & 32 & 0.53 & 0.24 \\
\textbf{Instance 4} & 23 & 43 & 12 & 11 & 31 & 0.52 & 0.28 \\
\textbf{Instance 5} & 31 & 48 & 17 & 14 & 31 & 0.55 & 0.35 \\
\textbf{Instance 6} & 33 & 49 & 18 & 15 & 31 & 0.55 & 0.37 \\
\textbf{Instance 7} & 37 & 49 & 19 & 18 & 30 & 0.51 & 0.39 \\
\textbf{Instance  8} & 40 & 49 & 20 & 20 & 29 & 0.5 & 0.41 \\
\textbf{Instance  9} & 42 & 49 & 20 & 22 & 29 & 0.48 & 0.41 \\
\textbf{Instance 10} & 43 & 50 & 20 & 23 & 30 & 0.47 & 0.4 \\
\bottomrule
\end{tabular}%
}
\end{table}

\begin{table}[]
\centering
\caption{RQ2 Webex results}\label{table:RQ2 Webex}
\resizebox{\columnwidth}{!}{%
\begin{tabular}{llllllll}
\toprule
\multicolumn{8}{c}{\textbf{Webex}} \\
\midrule
\textbf{Instance \#} & \textbf{\begin{tabular}[c]{@{}l@{}}Hark \\ issues\end{tabular}} & \textbf{\begin{tabular}[c]{@{}l@{}}Pre-PI \\ issues\end{tabular}} & \textbf{\begin{tabular}[c]{@{}l@{}}Common \\ issues\end{tabular}} & \textbf{\begin{tabular}[c]{@{}l@{}}Unique \\ Hark \\ issues\end{tabular}} & \textbf{\begin{tabular}[c]{@{}l@{}}Unique \\ Pre-PI \\ issues\end{tabular}} & \textbf{\begin{tabular}[c]{@{}l@{}}Hark \\ overlap \\ ratio\end{tabular}} & \textbf{\begin{tabular}[c]{@{}l@{}}Pre-PI \\ overlap \\ ratio\end{tabular}} \\
\midrule
\textbf{Instance 1} & 14 & 30 & 9 & 5 & 21 & 0.64 & 0.3 \\
\textbf{Instance 2} & 24 & 30 & 16 & 8 & 14 & 0.67 & 0.53 \\
\textbf{Instance 3} & 30 & 31 & 17 & 13 & 14 & 0.57 & 0.55 \\
\textbf{Instance 4} & 36 & 31 & 18 & 18 & 13 & 0.5 & 0.58 \\
\textbf{Instance 5} & 43 & 35 & 20 & 23 & 15 & 0.47 & 0.57 \\
\textbf{Instance 6} & 51 & 39 & 22 & 29 & 17 & 0.43 & 0.56 \\
\textbf{Instance 7} & 54 & 39 & 25 & 29 & 14 & 0.46 & 0.64 \\
\textbf{Instance 8} & 57 & 39 & 25 & 32 & 14 & 0.44 & 0.64 \\
\textbf{Instance 9} & 60 & 43 & 26 & 34 & 17 & 0.43 & 0.6 \\
\textbf{Instance 10} & 60 & 43 & 26 & 34 & 17 & 0.43 & 0.6 \\
\textbf{Instance 11} & 60 & 43 & 26 & 34 & 17 & 0.43 & 0.6 \\
\textbf{Instance 12} & 61 & 43 & 27 & 34 & 16 & 0.44 & 0.63 \\
\textbf{Instance 13} & 62 & 44 & 28 & 34 & 16 & 0.45 & 0.64 \\
\bottomrule
\end{tabular}%
}
\end{table} 

Next, we examine the overlap ratios, where we see some variation among the apps. In the Word app, Pre-PI initially struggles with a lower overlap ratio compared to baseline, but its performance improves over successive instances. For Zoom, the Pre-PI overlap ratios improve over successive instances but are lower compared to Hark overlap ratios. In contrast, Webex results clearly favor Pre-PI. Apart from the first few instances, Pre-PI consistently achieves higher overlap ratios compared to Hark in Webex. 

A lower overlap score for Pre-PI compared to baseline in the initial instances is not surprising, Pre-PI is able to identify a significantly higher number of issues even as early as instance 1, while Hark captures lower number of issues in the initial instances. This imbalance reduces the number of overlaps and initially hurts the performance of Pre-PI. However, as the releases progress, many of the issues identified by Pre-PI early appear later in Hark, leading to improvement in Pre-PI overlap scores. Motivated by these observations, we further investigate in RQ3 whether the issues identified by Pre-PI in earlier releases emerge in the later instances of Hark issues. 

Overall, these RQ2 results show that Pre-PI achieves an early lead by generating a larger number of issues and achieving faster convergence showing its potential of capturing privacy concerns ahead of the baseline. Although the overlap scores of Pre-PI is lower in the initial instances, this can be explained by its ability to identify issues that appear in Hark only in later releases which we explore in RQ3 next.

\begin{figure*}[t]
\centering
    \includegraphics[width=\textwidth]{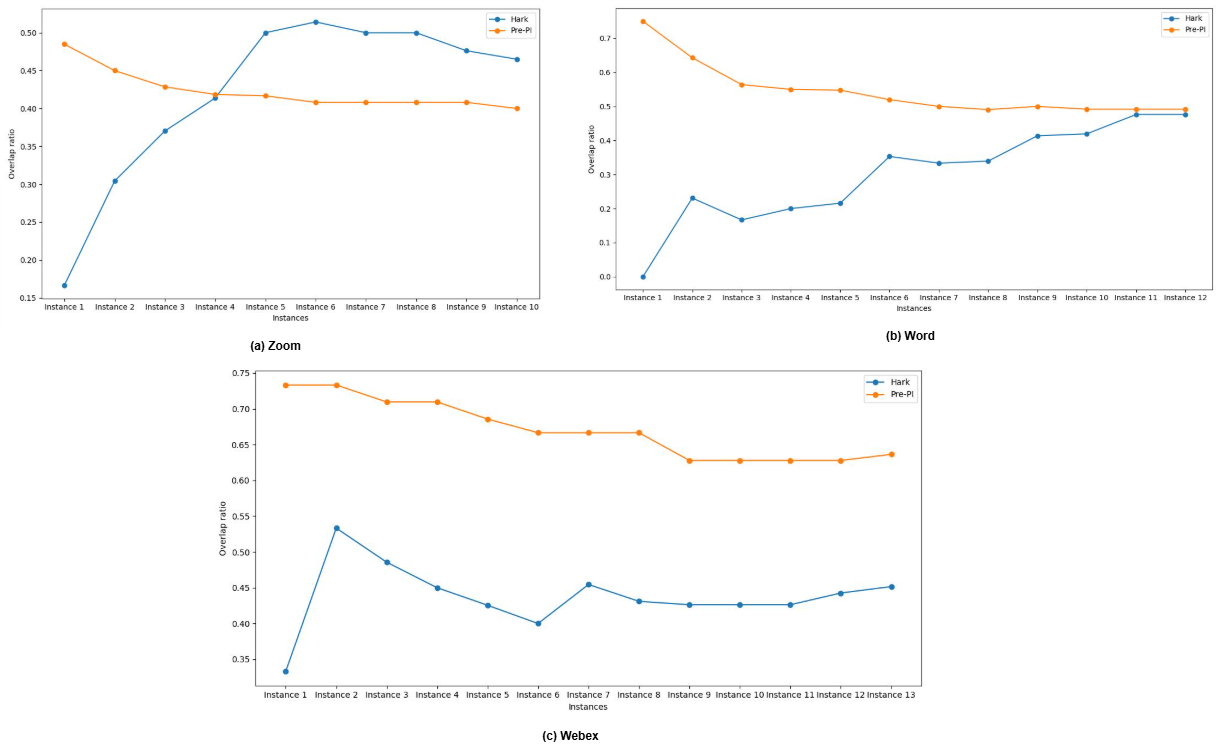}
    \caption{RQ3 results}
    \label{fig:rq3_combined}
\end{figure*}

To address RQ3, and to further determine the anticipatory power of Pre-PI, we study whether the issues generated by Pre-PI are later confirmed in future releases by the Hark baseline. To address this, we plot the temporal overlap ratios for Pre-PI and Hark across the three apps in Figure \ref{fig:rq3_combined}. 

For Webex, we observe that across all the instances the temporal overlap ratio for Pre-PI is significantly higher compared to baseline. In the Word app, we observe that Pre-PI achieves higher overlap ratios compared to baseline in the initial instances but the difference becomes less prominent over the later instances. In Zoom, Pre-PI shows a strong initial performance for the first few instances, with Hark overlap ratios slowly catching up and then achieving higher overlap ratios compared to Pre-PI in later stages. However, it is worth noting that across all the datasets, Pre-PI temporal overlap ratios remain relatively high and do not drop sharply. On the other hand, Hark temporal overlap ratio fluctuates and is significantly low in the initial instances with noticeable improvements only in some cases over later releases. 

Another trend across the datasets is that the performance of Pre-PI tends to decline over successive instances, whereas the performance of Hark generally improves with the exception of the Webex app. This trend can be attributed to the temporal aspect of evaluation. The overlapping issues are measured only up to the final release instance of the evaluation window (n), beyond which Hark issues are not available to validate further predictions. As a result, Pre-PI issues that may be identified by Hark as valid issues beyond the evaluation window are missed. This may falsely reduce the Pre-PI performance especially towards the final releases because of the evaluation window constraint. This could be attributed as the reason why Pre-PI performance drops over instances and baseline improves over the instances as the number of instances in the evaluation window reduces in size. Evaluating the performance with a sliding evaluation window is left as future work. 

Overall, from these results, we conclude that Pre-PI outperforms Hark and acts as an early privacy issues detection system, identifying privacy issues that the baseline Hark detects only after the users encounter and report them through user reviews. RQ3 highlights the value of Pre-PI with the temporal advantage, allowing privacy issues to be surfaced earlier than the baseline. 

\begin{figure*}[t]
\centering
    \includegraphics[width=\textwidth]{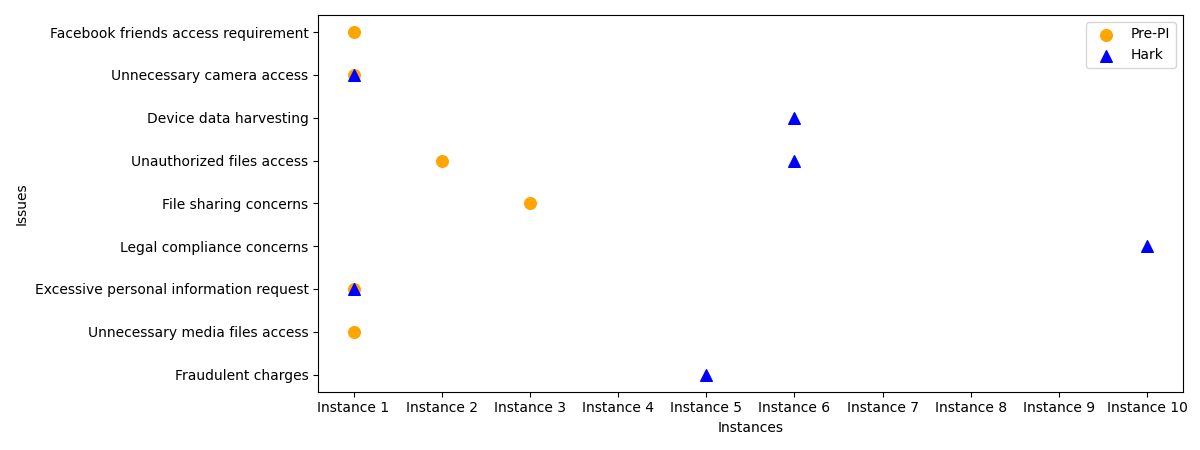}
    \caption{Examples of privacy issues across Word, Webex, and Zoom apps, showing the instances where each issue was first identified by Pre-PI and Hark.}
    \label{fig:illustrative_example}
\end{figure*}

We provide some examples of issues identified across different apps and categories in Table \ref{table:illustrative examples}. For each issue, we highlight whether it is common to both methods or unique to Pre-PI or Hark. Issues such as \textit{Unnecessary camera access} and \textit{Unauthorized files access} are common issues identified across both methods. However, Pre-PI also identifies additional feature-specific concerns such as \textit{Facebook friends access requirement} and \textit{Unnecessary media files access} that Hark fails to capture. Moreover, the issues that are unique to Hark such as \textit{Legal compliance concerns} and \textit{Fraudulent charges} are generic, broader, and less feature-specific feedback, highlighting Hark's reliance on actual user reviews. In Figure \ref{fig:illustrative_example}, for these same issues, we show the instances where the issue first appeared for Pre-PI and Hark. We observe from Figure \ref{fig:illustrative_example} that the general trend is Pre-PI detects issues in earlier release instances compared to Hark, demonstrating Pre-PI's advantage in anticipating privacy concerns using candidate features in the absence of user reviews. 

\begin{table}[]
\caption{Examples of privacy issues, along with their associated app and category.}\label{table:illustrative examples}
\begin{tabular}{@{}lll@{}}
\toprule
\multicolumn{1}{c}{\textbf{App}} & \multicolumn{1}{c}{\textbf{Issue}} & \multicolumn{1}{c}{\textbf{Category}} \\ \midrule
Webex & Unnecessary camera access & Common issue \\
Webex & Facebook friends access requirement & Unique Pre-PI issue \\
Webex & Device data harvesting & Unique Hark issue \\
Word & Unauthorized files access & Common issue \\
Word & File sharing concerns & Unique Pre-PI issue \\
Word & Legal compliance concerns & Unique Hark issue \\
Zoom & Excessive personal information request & Common issue \\
Zoom & Unnecessary media files access & Unique Pre-PI issue \\
Zoom & Fraudulent charges & Unique Hark issue \\ \bottomrule
\end{tabular}
\end{table}

\section{Threats to Validity}
In this section, we discuss the various threats to validity of our study. 

\textit{Threats to construct validity:} In our experiments, we assume that the privacy user reviews utilized by Hark to generate privacy issues are valid ground truth. However, user reviews may not capture all privacy concerns and may express them vaguely that could affect both the training and evaluation process. Also, the assumption that overlapping issues indicate correctness may be an oversimplification of the identification of privacy concerns problem.
Furthermore, we use synthetic reviews to generate issues that help simulate feedback for candidate features in the absence of actual user feedback. However, simulated feedback may differ from actual feedback and may not fully represent concerns that actual users are able to identify after the particular app version is released. 

\textit{Threats to internal validity:} The privacy classifier which is a multi-model ensemble may be too strict and focus on high precision, leaving out borderline privacy reviews. Although the classifier is motivated from prior work \cite{ICSE_22_google}, we may be selecting privacy reviews at the cost of recall. 
Our training and evaluation process could be impacted by the implementation choice. For example, for the T5 model, we utilized the parameter and hyperparameter settings from the previous literature; however, alternative parameter choices may yield different results. Therefore, we share the replication packet (https://doi.org/10.5281/zenodo.17274646) of our study that includes the datasets used and the code implementations to facilitate replication and expansion.

\textit{Threats to external validity:} Our evaluation is from three real-world apps that are widely used. However, the generalization of Pre-PI to other real-world domain apps is not explored. In addition, our release window and candidate feature selection is based on release notes. However, these release cycles and feedback loops may be different for software vendors, especially across non-mobile settings. Another external validity threat is from our choice of evaluators for RQ1. For the human subejct study we relied on long term users of the apps instead of actual developers of the apps. Although developers may provide stronger implementation insights, previous studies \cite{Nguyen_survey_2017_CCS, Acar_SP_16} have shown that hiring developers from Google Play is very difficult with a response rate of less than 1 percent. 

\section{Conclusion}
Mobile apps have become very personal, often requesting multiple forms of privacy-centric information in exchange for providing a more personalized user experience \cite{Ebrahmi_21_IST_review}. This has led to multiple researchers exploring privacy concerns for mobile apps \cite{Ebrahmi_21_IST_review}. User feedback from app stores provides software vendors and developers with a direct channel for collecting user input to guide the evolution of their apps. Privacy related feedback is particularly valuable, as unresolved concerns can make or break an app, determining its overall success. While much of the existing work focused on summarizing privacy concerns for the whole app from user feedback, we introduce a novel approach that uses user feedback to associate privacy concerns with software features and simulate feedback for candidate features of upcoming releases to identify potential privacy concerns. We introduce Pre-PI that uses state-of-the-art ML and NLP techniques. Our Pre-PI results indicate that it produces valid privacy concerns, aligns well with privacy concerns identified by baseline over releases, even without access to actual user feedback, and further is able to identify privacy concerns earlier than the baseline. 
In our future work, we will be experimenting on other apps from different domains, leverage sliding evaluation window, and explore LLM based approaches for simulating feedback.

\bibliographystyle{IEEEtran}
\bibliography{references}

\end{document}